# Modulational instability of coupled nonlinear field equations for pulse propagation in a negative index material embedded into a Kerr medium

**Amarendra K. Sarma\* and Manirupa Saha**
Department of Physics, Indian Institute of Technology Guwahati, Guwahati-781039, Assam, India.
\*Corresponding author: aksarma@iitg.ernet.in

We have investigated the modulational instability (MI) in a negative index media (NIM) using a new generalized model describing the pulse propagation in a negative index material embedded into a Kerr medium. We have found that one could control the gain of MI in a NIM by tuning the initial electric or magnetic field amplitudes. Our model successfully recovers previously proposed models to describe pulse propagation in NIMs exhibiting Kerr nonlinearity. Moreover it contains a few additional terms connecting both the electric and magnetic field envelopes in a NIM.

## 1. INTRODUCTION

Since the publication of seminal papers by Pendry [1] and Smith et al. [2], metamaterials or left-handed materials have become a topic of intense theoretical and experimental research. It has attracted researchers from various disciplines in science and engineering making it a interdisciplinary research subject in true sense. Metamaterials are man-made artificial structures which exhibit uncommon properties such as reversal of Snell's law, reversal of Doppler effect and Cherenkov effect, etc. [3]. One key feature of metamaterials is that they show negative refractive index, for which they are often referred to as negative index material (NIM), owing to their simultaneous negative electric permittivity and negative magnetic permeability. In this work, now onwards we would use the word NIM instead of metamaterial. The first possibility of NIM was theoretically introduced by Veselago in 1968 [4]. Because of the recent experimental realization of negative index materials at infrared and optical frequencies [5-7] metamaterial or NIM research is getting a new dimension. Various potential applications of NIMs have been proposed and studied [8]. In this context, the study of nonlinear pulse propagation, particularly, optical solitons is a new and extremely exciting field of research [9-11]. This may be mainly due to the fact, apart from the richness of physics, that metamaterials or NIMs being artificially structured materials, we might have the flexibility of controlling these pulses at our will. Many authors have investigated and proposed nonlinear pulse propagation model in NIMs in various contexts. The first significant attempt in deriving a proper model equation to describe nonlinear pulse propagation was made by Scalora et al. [12]. They derived a new generalized nonlinear Schrodinger equation, without taking nonlinear magnetization into account, describing the propagation of ultra short pulses in bulk negative index media exhibiting frequency dependent dielectric susceptibility and magnetic permeability. Following an almost similar approach, but eliminating the magnetic field at the very start, Wen et al. [13] have derived a (3+1) dimensional evolution equation for NIM. On the other hand, Lazarides and Tsironis [14] derived a system of coupled nonlinear Schrodinger equation (NLSE) for the envelopes of the propagating electric and magnetic fields in an isotropic and homogeneous nonlinear left-handed material taking both nonlinear polarization and magnetization into account. However, they used a perturbative approach and focused only on a completely integrable Manakov type model. Again in Ref. [15], Wen et al. have derived a coupled NLSE suitable for few cycle pulse propagation in a NIM with both nonlinear electric and magnetic polarization. More recently, using the reductive perturbative method, Tsitsas et al. [16] derived a higher order NLSE to describe pulse propagation in NIMs. In this paper, we derive a mathematical model describing pulse propagation in a NIM created out of split ring resonators and arrays of wires embedded in a Kerr medium, exhibiting both electric and magnetic nonlinearity of Kerr type. Unlike other authors we invoke the electric and magnetic Kerr effect quite early into the derivation. We successfully reproduce previously proposed models under appropriate approximations from our newly derived propagation equation. Our model predicts some new additional features as well. We have also offered a brief modulation instability analysis considering its significance in relation to the existence of optical solitons in negative index materials using the newly derived model equations.

## 2. THEORETICAL MODEL

We consider a nonlinear negative index material embedded in a Kerr medium characterized by the following form of nonlinear electric polarization ($\mathbf{P}_{NL}$) and nonlinear magnetization ($\mathbf{M}_{NL}$): $\mathbf{P}_{NL} = \varepsilon_{NL}\mathbf{E} = \varepsilon_0 \chi_P^{(3)} |\mathbf{E}|^2 \mathbf{E}, \mathbf{M}_{NL} = \mu_{NL}\mathbf{H} = \mu_0 \chi_M^{(3)} |\mathbf{H}|^2 \mathbf{H}$; where $\varepsilon_{NL}$ and $\mu_{NL}$ are respectively the nonlinear electric permittivity and nonlinear magnetic permeability while $\chi_P^{(3)}$, $\chi_M^{(3)}$ are the respective third order electric and magnetic susceptibility. $\mathbf{E}$ and $\mathbf{H}$ are the electric and magnetic fields respectively. The nonlinear pulse propagation through metamaterials is characterized by electric flux density ($\mathbf{D}$) and magnetic induction ($\mathbf{B}$) which depends on electric ($\mathbf{E}$) and magnetic($\mathbf{H}$) field intensities as $\mathbf{D} = \varepsilon\mathbf{E} + \mathbf{P}_{NL}$ and $\mathbf{B} = \mu\mathbf{H} + \mathbf{M}_{NL}$. The dielectric permittivity ($\varepsilon$) and magnetic permeability ($\mu$) are dispersive in NIMs otherwise the energy density could be negative and their frequency dispersion is given by the lossless Drude model [3]. In principle one must take loss into account but for the sake of simplicity, like other authors, we are ignoring it. We assume that both the electric and magnetic field is propagating in a uniform, bulk NIM containing no free charge and no free current. It is straightforward to get the following nonlinear pulse propagation equations from the Maxwell equations [15, 17]:

$$\left(\frac{\partial^2}{\partial z^2} + \nabla_\perp^2\right)\mathbf{E}(r,t) - \nabla\left(\nabla\cdot\mathbf{E}(r,t)\right) = \mu\varepsilon\frac{\partial^2\mathbf{E}(r,t)}{\partial t^2} + \mu\frac{\partial^2\mathbf{P}_{NL}(r,t)}{\partial t^2} + \frac{\partial}{\partial t}\left(\nabla\times\mathbf{M}_{NL}(r,t)\right)$$
$$\left(\frac{\partial^2}{\partial z^2} + \nabla_\perp^2\right)\mathbf{H}(r,t) - \nabla\left(\nabla\cdot\mathbf{H}(r,t)\right) = \mu\varepsilon\frac{\partial^2\mathbf{H}(r,t)}{\partial t^2} + \varepsilon\frac{\partial^2\mathbf{M}_{NL}(r,t)}{\partial t^2} - \frac{\partial}{\partial t}\left(\nabla\times\mathbf{P}_{NL}(r,t)\right)$$

(1)

In order to simplify the above complex equations we make a couple of approximations: Firstly we assume that both the electric and magnetic field is propagating along the z direction and are linearly polarized. Secondly, there are negligible transverse inhomogeneities of the medium polarization and magnetization. Under these approximations we obtain the following set of coupled equations for the electric and magnetic fields in a NIM embedded into a Kerr medium.

$$\left(\frac{\partial^2}{\partial z^2}+\nabla_\perp^2\right)\mathbf{E}(r,t)=\mu\varepsilon\frac{\partial^2\mathbf{E}(r,t)}{\partial t^2}+\varepsilon_0\chi_E^{(3)}\mu\frac{\partial^2}{\partial t^2}\left(|\mathbf{E}|^2\mathbf{E}(r,t)\right)+\mu_{NL}\varepsilon\frac{\partial^2\mathbf{E}(r,t)}{\partial t^2}$$
$$\left(\frac{\partial^2}{\partial z^2}+\nabla_\perp^2\right)\mathbf{H}(r,t)=\mu\varepsilon\frac{\partial^2\mathbf{H}(r,t)}{\partial t^2}+\mu_0\chi_M^{(3)}\varepsilon\frac{\partial^2}{\partial t^2}\left(|\mathbf{H}|^2\mathbf{H}(r,t)\right)+\varepsilon_{NL}\mu\frac{\partial^2\mathbf{H}(r,t)}{\partial t^2}$$
(2)

Using the Fourier transformations of the fields as $\mathbf{E}(r,t)=\frac{1}{2\pi}\int_{-\infty}^{+\infty}\tilde{\mathbf{E}}(r,\omega)e^{-i\omega t}d\omega$ and $\mathbf{H}(r,t)=\frac{1}{2\pi}\int_{-\infty}^{+\infty}\tilde{\mathbf{H}}(r,\omega)e^{-i\omega t}d\omega$, Eq. (2) could be expressed in the frequency domain as follows

$$\left(\frac{\partial^2}{\partial z^2}+\nabla_\perp^2\right)\tilde{\mathbf{E}}(r,\omega)=-\omega^2\mu(\omega)\varepsilon(\omega)\tilde{\mathbf{E}}(r,\omega)-\varepsilon_0\chi_E^{(3)}\omega^2\mu(\omega)|\tilde{\mathbf{E}}|^2\tilde{\mathbf{E}}(r,\omega)-\mu_{NL}\omega^2\varepsilon(\omega)\tilde{\mathbf{E}}(r,\omega)$$
$$\left(\frac{\partial^2}{\partial z^2}+\nabla_\perp^2\right)\tilde{\mathbf{H}}(r,\omega)=-\omega^2\mu(\omega)\varepsilon(\omega)\tilde{\mathbf{H}}(r,\omega)-\mu_0\chi_M^{(3)}\omega^2\varepsilon(\omega)|\tilde{\mathbf{H}}|^2\tilde{\mathbf{H}}(r,\omega)-\varepsilon_{NL}\omega^2\mu(\omega)\tilde{\mathbf{H}}(r,\omega)$$
(3)

Our aim is, however, to obtain a set of coupled equations for the envelopes of the electric and magnetic fields. So we introduce the envelope and carrier form for the electric field and magnetic field in the usual way:

$$\mathbf{E}(r,t)=\hat{x}A(r,t)\exp(ik_0z-i\omega_0t)+c.c$$
$$\mathbf{H}(r,t)=\hat{y}B(r,t)\exp(ik_0z-i\omega_0t)+c.c$$
(4)

where c.c. is the complex conjugate of the pulse. Here A and B are the slowly varying pulse envelopes of electric and magnetic field respectively. $k_0=\omega_0 n(\omega_0)/c$ is the wave number at the central frequency of the electromagnetic pulse and $n(\omega_0)$ is the refractive index of the material at $\omega_0$. Now taking the inverse Fourier transformation of Eq. (3) and then substituting Eq.(4), we obtain the following coupled equations for the envelopes of the electric and magnetic fields

$$\frac{\partial^2 A}{\partial z^2}+2ik_0\frac{\partial A}{\partial z}-k_0^2A+\nabla_\perp^2A=-\sum_{m=0}^\infty D_m\frac{\partial^m A}{\partial t^m}-\varepsilon_0\chi_E^{(3)}\left(\omega_0+i\frac{\partial}{\partial t}\right)\sum_{m=0}^\infty F_m\frac{\partial^m}{\partial t^m}|A|^2A-\mu_0\chi_M^{(3)}|B|^2\left(\omega_0+i\frac{\partial}{\partial t}\right)\sum_{m=0}^\infty G_m\frac{\partial^m A}{\partial t^m}$$
$$\frac{\partial^2 B}{\partial z^2}+2ik_0\frac{\partial B}{\partial z}-k_0^2B+\nabla_\perp^2B=-\sum_{m=0}^\infty D_m\frac{\partial^m B}{\partial t^m}-\mu_0\chi_M^{(3)}\left(\omega_0+i\frac{\partial}{\partial t}\right)\sum_{m=0}^\infty G_m\frac{\partial^m}{\partial t^m}|B|^2B-\varepsilon_0\chi_E^{(3)}|A|^2\left(\omega_0+i\frac{\partial}{\partial t}\right)\sum_{m=0}^\infty F_m\frac{\partial^m B}{\partial t^m}$$
(5)

where

$$D_m = \sum_{l=0}^{m} \frac{i^m}{l!(m-l)!} \frac{\partial^l(\omega\varepsilon)}{\partial\omega^l}\bigg|_{\omega=\omega_0} \frac{\partial^{m-l}(\omega\mu)}{\partial\omega^{m-l}}\bigg|_{\omega=\omega_0}, \quad F_m = \frac{i^m}{m!}\frac{\partial^m(\omega\mu)}{\partial\omega^m}\bigg|_{\omega=\omega_0} \text{ and } G_n = \frac{i^m}{m!}\frac{\partial^l(\omega\varepsilon)}{\partial\omega^l}\bigg|_{\omega=\omega_0} \quad (6)$$

Now we introduce the travelling coordinates: $\xi = z, \tau = t - \beta_1 z$. Keeping the linear dispersion terms up to second order and nonlinear dispersion terms up to first order, Eq.(5) could be reduced to the following form:

$$\frac{\partial A}{\partial \xi} = \frac{i}{2k_0}\nabla_\perp^2 A - \frac{i\beta_2}{2}\frac{\partial^2 A}{\partial \tau^2} + \frac{1}{ik_0 V}\frac{\partial^2 A}{\partial \xi \partial \tau} + \frac{i\omega_0^2 \mu(\omega_0)\chi_E^{(3)}}{2k_0}|A|^2 A - \frac{\omega_0^2 \mu(\omega_0)\varepsilon_0 \chi_E^{(3)}}{2k_0}\frac{1}{\omega_0}\left(1 + \frac{\gamma}{\mu(\omega_0)}\right)\frac{\partial |A|^2 A}{\partial \tau}$$

$$+ \frac{i\omega_0^2 \varepsilon(\omega_0)\mu_0 \chi_M^{(3)}}{2k_0}|B|^2 A - \frac{\mu_0 \chi_M^{(3)}\omega_0 \varepsilon(\omega_0)}{2k_0}|B|^2\left(1 + \frac{\alpha}{\varepsilon(\omega_0)}\right)\frac{\partial A}{\partial \tau}$$

$$\frac{\partial B}{\partial \xi} = \frac{i}{2k_0}\nabla_\perp^2 B - \frac{i\beta_2}{2}\frac{\partial^2 B}{\partial \tau^2} + \frac{1}{ik_0 V}\frac{\partial^2 B}{\partial \xi \partial \tau} + \frac{i\omega_0^2 \varepsilon(\omega_0)\chi_M^{(3)}}{2k_0}|B|^2 B - \frac{\omega_0^2 \varepsilon(\omega_0)\varepsilon_0 \chi_M^{(3)}}{2k_0}\frac{1}{\omega_0}\left(1 + \frac{\alpha}{\varepsilon(\omega_0)}\right)\frac{\partial |B|^2 B}{\partial \tau}$$

$$+ \frac{i\omega_0^2 \mu(\omega_0)\varepsilon_0 \chi_E^{(3)}}{2k_0}|A|^2 B - \frac{\varepsilon_0 \chi_E^{(3)}\omega_0 \varepsilon(\omega_0)}{2k_0}|A|^2\left(1 + \frac{\gamma}{\mu(\omega_0)}\right)\frac{\partial B}{\partial \tau}$$

(7)

where $\beta_2 = \left[\{\alpha\gamma + \omega_0 \mu(\omega_0)\alpha'/2 + \omega_0 \varepsilon(\omega_0)\gamma'/2 - 1/V^2\}/k_0\right]$ with $\gamma = \partial[\omega\mu(\omega)]/\partial\omega\big|_{\omega=\omega_0}$, $\gamma' = \partial^2[\omega\mu(\omega)]/\partial^2\omega\big|_{\omega=\omega_0}$, $\alpha = \partial[\omega\varepsilon(\omega)]/\partial\omega\big|_{\omega=\omega_0}$, $\alpha' = \partial^2[\omega\varepsilon(\omega)]/\partial^2\omega\big|_{\omega=\omega_0}$ and $V = 2k_0/[\omega_0 \varepsilon(\omega_0)\gamma + \omega_0 \mu(\omega_0)\alpha]$.

In order to make the above propagation model applicable and solvable we make some further approximations [12]: $\frac{\partial^2 A}{\partial \xi \partial \tau} = \frac{i\omega_0^2 \mu(\omega_0)\chi_E^{(3)}}{2k_0}\frac{\partial |A|^2 A}{\partial \tau}$ and $\frac{\partial^2 B}{\partial \xi \partial \tau} = \frac{i\omega_0^2 \varepsilon(\omega_0)\chi_M^{(3)}}{2k_0}\frac{\partial |B|^2 B}{\partial \tau}$.

Under these assumptions we obtain the following coupled generalized NLSE for a nonlinear negative index material exhibiting Kerr type electric and magnetic nonlinear polarization:

$$\frac{\partial A}{\partial \xi} = \frac{i}{2k_0}\nabla_\perp^2 A - \frac{i\beta_2}{2}\frac{\partial^2 A}{\partial \tau^2} + iP_{nl}\left(1 + iP_s \frac{\partial}{\partial \tau}\right)|A|^2 A + iQ_{nl}|B|^2\left(A + iP_{se}\frac{\partial A}{\partial \tau}\right)$$

$$\frac{\partial B}{\partial \xi} = \frac{i}{2k_0}\nabla_\perp^2 B - \frac{i\beta_2}{2}\frac{\partial^2 B}{\partial \tau^2} + iQ_{nl}\left(1 + iQ_s \frac{\partial}{\partial \tau}\right)|B|^2 B + iP_{nl}|A|^2\left(B + iQ_{sh}\frac{\partial B}{\partial \tau}\right)$$

(8)

where

$$P_{nl} = \frac{\omega_0^2 \mu(\omega_0)\varepsilon_0 \chi_E^{(3)}}{2k_0}; \quad P_s = \left[\frac{1}{\omega_0}\left(1+\frac{\gamma}{\mu(\omega_0)}\right) - \frac{1}{k_0 V}\right]; \quad P_{se} = \frac{1}{\omega_0}\left(1+\frac{\alpha}{\varepsilon(\omega_0)}\right)$$

$$Q_{nl} = \frac{\omega_0^2 \varepsilon(\omega_0)\mu_0 \chi_M^{(3)}}{2k_0}; \quad Q_s = \left[\frac{1}{\omega_0}\left(1+\frac{\alpha}{\varepsilon(\omega_0)}\right) - \frac{1}{k_0 V}\right]; \quad Q_{sh} = \frac{1}{\omega_0}\left(1+\frac{\gamma}{\mu(\omega_0)}\right)$$

(9)

$P_{nl}$ and $P_s$ are respectively the nonlinear and self-steepening coefficients for the electric field. We name $P_{se}$ as the electric coupling coefficient. $Q_{nl}$ and $Q_s$ are the corresponding coefficients for the magnetic field, while we call $Q_{sh}$ as the magnetic coupling coefficient. Eq. (8) is the generalized coupled NLSE for pulse propagation for a negative index material embedded into a Kerr medium, which is one of the main results of this work. It should be noted that if $Q_{nl} = 0$ we recover exactly the same equation in Ref. [13] for the envelope of the electric field. However our model contains a few additional terms compared to previous models. For example, the last two terms of Eq. (8) connecting both the electric and magnetic field envelopes in a NIM. In Fig. 1(a) we plot the variation of $n, P_{nl}, P_s$ and $P_{se}$ with the normalized frequency $\omega_0/\omega_{pe}$ for $\omega_{pm}/\omega_{pe} = 0.8$ while in Fig.1(b) we plot the corresponding variations for $Q_{nl}, Q_s$ and $Q_{sh}$. Here $\omega_{pe}$ and $\omega_{pm}$ are the respective electric and magnetic plasma frequency and the parameters are plotted at $\omega = \omega_0$.

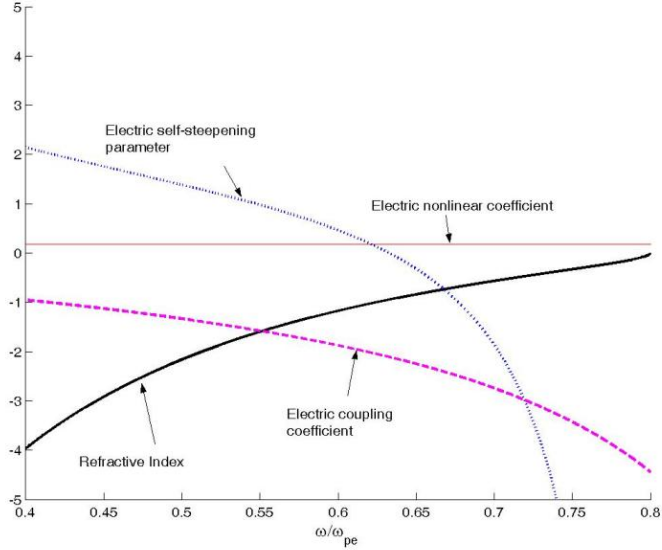

Fig.1 (a) (Color online) Variation of refractive index n, electric nonlinear coefficient $P_{nl}$, electric self-steepening parameter $P_s$ and electric coupling coefficient $P_{se}$ with the normalized frequency $\omega_0/\omega_{pe}$ with $\omega_{pm}/\omega_{pe} = 0.8$. $P_{nl}$ is calculated in units of $\omega_{pe}\chi_E^{(3)}/c$ while $P_s$ and $P_{se}$ are calculated in the units of $1/\omega_{pe}$.

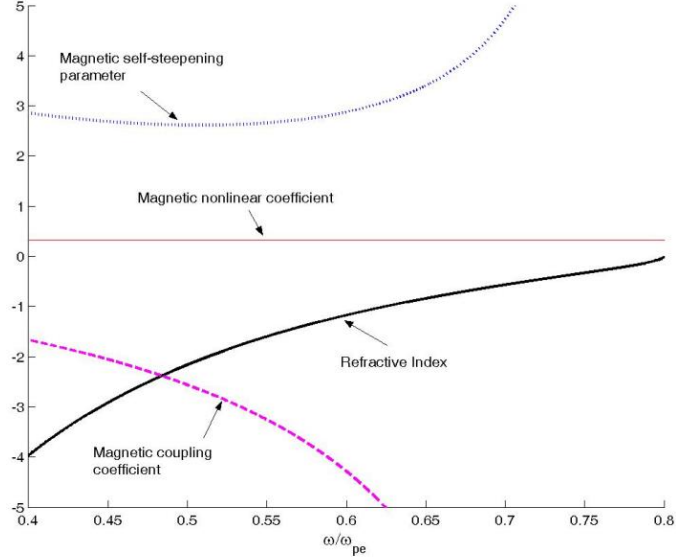

Fig.1 (b) (Color online) Variation of refractive index n, magnetic nonlinear coefficient $Q_{nl}$, magnetic self-steepening parameter $Q_s$ and magnetic coupling coefficient $Q_{sh}$ with the normalized frequency $\omega_0/\omega_{pe}$ with $\omega_{pm}/\omega_{pe}=0.8$. $Q_{nl}$ is calculated in units of $\omega_{pe}\chi_M^{(3)}/c$ while $Q_s$ and $Q_{sh}$ are calculated in the units of $1/\omega_{pe}$.

For simplicity of calculations it is convenient to write the coupled field equations in normalized units. First we assume the normalized variables as [15]:

$$Z=\frac{\xi}{L_D}, T=\frac{\tau}{T_0}; U=\frac{A}{A_0}, V=\frac{B}{B_0}; X=\frac{x}{L_\perp}, Y=\frac{y}{L_\perp} \qquad (10)$$
$$u=N_E U, v=N_H V$$

where $T_0$ is the pulse width, $L_D=T_0^2/|\beta_2|$ is the dispersion length and $A_0$ and $B_0$ are the initial amplitude of electric and magnetic field. $N_E$ and $N_H$ may be termed as the order of soliton for the electric and the magnetic field, defined as $N_E^2=L_D/L_{Pnl}, N_H^2=L_D/L_{Mnl}$. In this work we assume $N_E=N_H=N$. We define the nonlinear polarization length as $L_{Pnl}=1/P_{nl}A_0^2$ while the nonlinear magnetization length as $L_{Mnl}=1/Q_{nl}B_0^2$. A characteristic length $L_\perp=\sqrt{|L_D/k_0|}$ is also introduced. The above equation (8) is thus transformed to the following normalized form:

$$\frac{\partial u}{\partial Z}=\frac{i\,\mathrm{sgn}(k_0)}{2}\nabla_T^2 u-\frac{i\,\mathrm{sgn}(\beta_2)}{2}\frac{\partial^2 u}{\partial T^2}+i\left(1+iS_E\frac{\partial}{\partial T}\right)|u|^2 u+i|v|^2 u-C_E|v|^2\frac{\partial u}{\partial T}$$
$$\frac{\partial v}{\partial Z}=\frac{i\,\mathrm{sgn}(k_0)}{2}\nabla_T^2 v-\frac{i\,\mathrm{sgn}(\beta_2)}{2}\frac{\partial^2 v}{\partial T^2}+i\left(1+iS_H\frac{\partial}{\partial T}\right)|v|^2 v+i|u|^2 v-C_H|u|^2\frac{\partial v}{\partial T} \qquad (11)$$

where $\nabla_\perp^2 = \partial^2/\partial X^2 + \partial^2/\partial Y^2$ is the transverse Laplacian, $S_E = P_s/T_0$ is the electric self-steepening parameter, $S_H = Q_s/T_0$ is the magnetic self-steepening parameter, $C_E = P_{se}/T_0$ is electric coupling coefficient and $C_H = Q_{sh}/T_0$ is the magnetic coupling coefficient in normalized units.

## 3. MODULATIONAL INSTABILITY ANALYSIS

Now we would carry out a modulational instability (MI) analysis of the above generalized coupled field equations mainly due to its close relation with the existence of solitary waves or solitons. It is well known that modulational instability is a fundamental and ubiquitous process that appears in most nonlinear systems in nature. It occurs as a result of interplay between the nonlinearity and dispersion in time domain or diffraction in spatial domain. Many authors have done modulation instability analysis in negative index material in various contexts [13, 18-20]. In this work we are mainly interested in knowing the role of both the electric and the magnetic self-steepening parameters on MI. So we neglect the diffraction and the last term of the generalized model. It should be noted that Wen et al. [17] have already investigated MI in NIM to understand the role of self-steepening effect. But unlike them we are considering both the electric and the magnetic self-steepening effects. Moreover our model contains a term relating both the electric and magnetic field envelopes:

$$\frac{\partial u}{\partial Z} = -\frac{i\,\text{sgn}(\beta_2)}{2}\frac{\partial^2 u}{\partial T^2} + i\left(1 + iS_E\frac{\partial}{\partial T}\right)|u|^2 u + i|v|^2 u$$

$$\frac{\partial v}{\partial Z} = -\frac{i\,\text{sgn}(\beta_2)}{2}\frac{\partial^2 v}{\partial T^2} + i\left(1 + iS_H\frac{\partial}{\partial T}\right)|v|^2 v + i|u|^2 v$$

(12)

We assume that Eq.(12) have two steady state continuous wave solutions $u = a_0\exp(i\Omega_{0a}Z)$ and $v = b_0\exp(i\Omega_{0b}Z)$ where $a_0$ and $b_0$ are the normalized amplitude of electric and magnetic field envelopes and $\Omega_{0a}$ and $\Omega_{0b}$ are the corresponding nonlinear phase-shifts. Here, $\Omega_{0a} = \Omega_{0b} = a_0^2 + b_0^2$. Now let the CW solutions be slightly perturbed from the steady state such that

$$A(Z,T) = [a_0 + a(Z,T)]\exp(i\Omega_{0a}T) \text{ and } B(Z,T) = [b_0 + b(Z,T)]\exp(i\Omega_{0b}Z)$$

(13)

where $a(Z,T)$ and $b(Z,T)$ are the perturbations such that $a,b \ll 1$.

Substituting Eq. (13) in Eq. (12) and then linearizing in $a$ and $b$ we obtain the following evolution equations for the perturbations:

$$\frac{\partial a}{\partial Z} = -\frac{i\,\text{sgn}(\beta_2)}{2}\frac{\partial^2 a}{\partial T^2} + ia_0^2(a+a^*) + ia_0b_0(b+b^*) - S_E a_0^2\left(2\frac{\partial a}{\partial \tau} + \frac{\partial a^*}{\partial \tau}\right)$$
$$\frac{\partial b}{\partial Z} = -\frac{i\,\text{sgn}(\beta_2)}{2}\frac{\partial^2 b}{\partial T^2} + ib_0^2(b+b^*) + ia_0b_0(a+a^*) - S_H b_0^2\left(2\frac{\partial b}{\partial \tau} + \frac{\partial b^*}{\partial \tau}\right)$$
(14)

Separating the perturbation to real and imaginary parts, according to $a = a_1 + ia_2$, $b = b_1 + ib_2$ and assuming $(a_i, b_i) = (a_{i0}, b_{i0})\exp[i(KZ - \Omega T)]$, (for $i = 1, 2$) where $K$ and $\Omega$ are the wave number and the frequency of perturbation respectively in normalized units, from Eq.(14) we obtain the following relevant dispersion relation:

$$K = \frac{1}{2}\left[4s\Omega \pm \left(4s^2\Omega^2 + \frac{\Omega^4}{4} + 4\delta(a_0^2 + b_0^2)\Omega^2\right)^{\frac{1}{2}}\right]$$
(15)

where $\delta = \text{sgn}(\beta_2)$. Here we assume that $S_E a_0^2 = S_H b_0^2 = s$ and take the frequency regime to be: $0.4 \leq \omega/\omega_{pe} \leq 0.6$. Now onwards we would term $s$ as the reduced self-steepening parameter. Because $S_E \neq S_H$, the initial normalized amplitudes of electric and magnetic fields would always have to be different, i.e. $a_0 \neq b_0$. The steady-state solution becomes unstable only when K has an imaginary part since perturbation then grows exponentially. It can be clearly seen from Eq. (15) that K becomes imaginary only in the anomalous dispersion regimes i.e. when $\delta = -1$ and if the following condition is satisfied: $(a_0^2 + b_0^2) > (s^2 + \Omega^2/16)$. Under these MI conditions we obtain the gain spectrum $g(\Omega)$ of the modulation instability as follows:

$$g(\Omega) = 2\,\text{Im}(K) = \left[4(a_0^2 + b_0^2)\Omega^2 - 4s^2\Omega^2 - \Omega^4/4\right]^{\frac{1}{2}}$$
(16)

Clearly, the MI gain spectrum depends on the initial amplitude of the electric and the magnetic field, the perturbation frequency and the electric and magnetic self-steepening parameters through $s$. The gain becomes maximum at two frequencies given by $\Omega_{max} = \pm 2\left[2(a_0^2 + b_0^2) - 2s^2\right]^{1/2}$ with a peak value $4(a_0^2 + b_0^2 - s^2)^{1/2}$. In Fig. 2 we depict the MI gain as a function of the normalized perturbation frequency for various values of $\omega/\omega_{pe}$ for a pulse width $T_0 = 10\,\text{fs}$ and $a_0 = 1$. We observe that the MI gain spectrum is symmetric with respect to $\Omega = 0$ and MI gain decreases with increase in the normalized frequency.

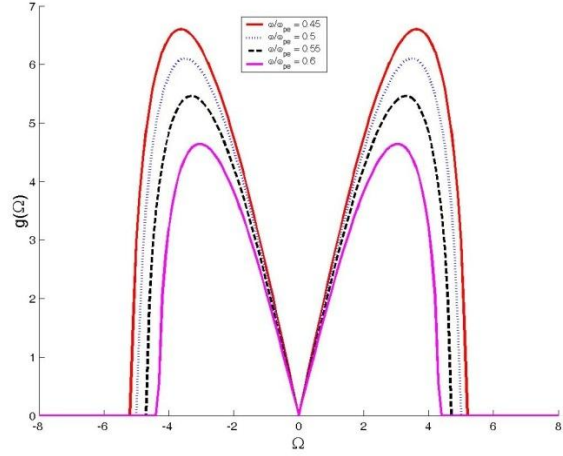

Fig. 2 (Color online) Modulational instability gain as a function of normalized perturbation frequency for four different values of $\omega/\omega_{pe}$ with $a_0=1$ and $T_0=10\,\text{fs}$.

This might be attributed to the constraint $S_E a_0^2 = S_H b_0^2 = s$ and the fact that with increase in the normalized frequency $\omega/\omega_{pe}$, as evident from Fig. 1, $S_E$ decreases while $S_H$ increases for a given value of $a_0$ and pulse width $T_0$ resulting in the reduction of both $s$ and $b_0$. The so called reduced self-steepening parameter $s$ could be controlled just by tuning the initial electric or magnetic field amplitudes for a given operating frequency of the NIM. In order to have a clear idea about the role of $s$ on MI, in Fig. 3 we plot the variation of MI gain as a function of the normalized perturbation frequency for various values of $s$ for a pulse with $T_0=10\,\text{fs}$ and $\omega/\omega_{pe}=0.5$. It may be noted that the corresponding values of the parameters $S_E$ and $S_H$ are 0.057 and 0.10 respectively.

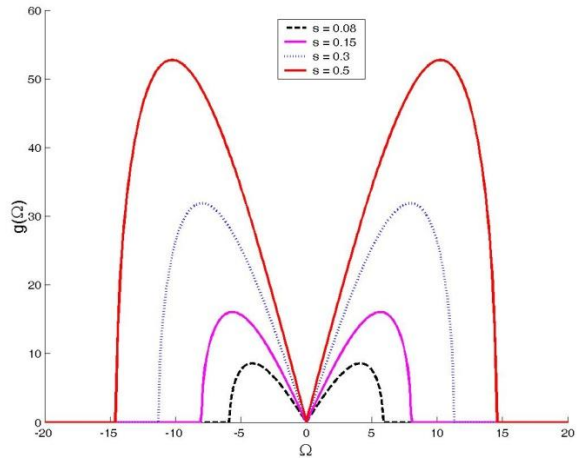

Fig. 3 (Color online) Modulational instability gain as a function of normalized perturbation frequency for four different values of reduced self-steepening parameter $s$ with $\omega/\omega_{pe}=0.5$ and $T_0=10\,\text{fs}$.

It can be clearly seen that with increase in $s$ the MI gain also increases, which should be due to the increase in the initial electric and magnetic field amplitudes. In passing we would like to mention that our study shows that, variation of the pulse width $T_0$ has negligibly small effect on MI gain spectrum. From the above analysis it is clear that by tuning the initial electric or magnetic field amplitudes or intensity one can control the modulational instability of electromagnetic pulses in a negative index media and thereby eventually could manipulate formation of solitons in it.

## 4. CONCLUSIONS

We have derived a generalized coupled nonlinear field equations for pulse propagation in a negative index material embedded into a Kerr medium. Our model successfully recovers previously proposed models to describe pulse propagation in NIMs exhibiting Kerr nonlinearity. Moreover it contains a few additional terms connecting both the electric and magnetic field envelopes in a NIM. We have also investigated the modulational instability in NIM using this new model under some restrictive conditions. We have found that one could control the gain of MI in a NIM by tuning the initial electric or magnetic field amplitudes of the pulse.

## ACKNOWLEDGEMENT

M. Saha would like to thank MHRD, Government of India for a research fellowship.

## REFERENCES

1. J. B. Pendry, "Negative refraction makes a perfect lens", Phys. Rev. Lett. **85**, 3966-3969 (2000).
2. D. R. Smith, W. J. Padilla, D. C. Vier, S. C. Nemat-Nasser, S. Schultz S, " Composite medium with simultaneously negative permeability and permittivity", Phys. Rev. Lett. **84**, 4184–4187 (2000).
3. C. Caloz and T. Itoh, *Electromagnetic Metamaterials: Transmission line theory and Microwave applications* (Wiley Interscience, 2006).
4. V. G. Veselago, "The electrodynamics of substance with simultaneously negative value of $\varepsilon$ and $\mu$," Sov. Phys. Usp. **10**, 509–514 (1968).
5. A. Berrier, M. Mulot, M. Swillo, M. Qiu, L. Thyl´en, A. Talneau and S. Anand, "Negative Refraction at Infrared-Wavelengths in a Two-Dimensional Photonic Crystal," Phys. Rev. Lett. **93**, 073902 (2004).
6. E. Schonbrun, M. Tinker, W. Park and J. -B. Lee, "Negative refraction in a Si-polymer photonic Crystal membrane," IEEE Photon. Technol. Lett. **17**, 1196- 1198, (2005).
7. V. M. Shalaev, W. Cai, U. K. Chettiar, H. Yuan, A. K. Sarychev, V. P. Drachev, and A. V. Kildishev, "Negative index of refraction in optical metamaterials," Opt. Lett. **30**, 3356-3358, (2005).
8. S. A. Ramakrishna and T. M. Grzegorczyk, *Physics and Applications of Negative refractive index materials* (CRC Press, 2009).
9. Liu, Y. M. and G. Bartal,"Subwavelength discrete solitons in nonlinear metamaterials," Phys.Rev. Lett. **99**, 153901(2007).
10. D'Aguanno, G., N. Mattiucci, M. Scalora, and M. J. Bloemer,"Bright and dark gap solitons in a negative index Fabry-Perot etalon," Phys. Rev. Lett. **93**, 213902-213905 (2004).

11. M. Scalora, D. D'Ceglia, G. D'Aguanno, N. Mattiucci, N. Akozbek, M. Centini, and M. J. Bloemer "Gap solitons in a nonlinear quadratic negative-index cavity," Phys. Rev. E **75**, 066606 (2007).
12. M. Scalora, M. S. Syrchin, N. Akozbek, E. Y. Poliakov, G. D'Aguanno, N. Mattiucci,M. J. Bloemer, and A. M. Zheltikov, "Generalized nonlinear Schrodinger equation for dispersive susceptibility and permeability: Application to negative index materials," Phys. Rev. Lett. **95**, 013902(2005).
13. S. Wen, Y. Wang, W. Su, Y. Xiang, and X. Fu, "Modulation instability in nonlinear negative-index material," Phys. Rev. E **73**, 036617 (2005).
14. N. Lazarides and G. P. Tsironis, "Coupled nonlinear Schrödinger field equations for electromagnetic wave propagation in nonlinear left-handed materials," Phys. Rev. E **71**, 036614 (2005).
15. S. Wen, Y. Xiang, X. Dai, Z. Tang, W. Su, and D. Fan, "Theoretical models for ultrashort electromagnetic pulse propagation in nonlinear metamaterials," Phys. Rev. A **75**,033815 (2007).
16. N. L. Tsitsas, N. Rompotis, I. Kourakis, P. G. Kevrekidis and D. J. Frantzeskakis, "Higher-order effects and ultrashort solitons in left-handed metamaterials", Phys. Rev. E **79**, 037601 (2009).
17. It should be noted that there may be other ways to do the analysis as pointed by the works of P. Kinsler: "Optical pulse propagation with minimal approximations", Phys. Rev. A 81, 013819 (2010) and "Unidirectional optical pulse propagation equation for materials with both electric and magnetic responses", Phys. Rev. A 81, 023808 (2010).
18. S. Wen, Y. Xiang, W. Su, Y. Hu, X. Fu, and D. Fan, "Role of the anomalous self-steepening effect in modulation instability in negative-index material," Opt. Express **14**, 1568–1575 (2006).
19. X. Dai, Y. Xiang, S. Wen and D. Fan,"Modulation instability of copropagating light beams in nonlinear metamaterials", J. Opt. Soc. Am. B **26**, 564–571 (2009).
20. I. Kourakis and P. K. Shukla, "Nonlinear propagation of electromagnetic waves in negative-refraction-index composite materials," Phys. Rev. E **72**, 016626 (2005).